\documentclass[12pt]{iopart}

\usepackage{iopams}  
\usepackage{graphicx}
\usepackage{array}
\usepackage{enumitem}
\usepackage{bm}% bold math
\usepackage[english,activeacute]{babel}

\begin{document}
\pagestyle{plain}
\selectlanguage{english}
%-----header----
\title[Discarded particles in shower simulations: effect on Missing and Invisible Energy]{Discarded low energy particles in extensive air shower simulations: Effect on the shower Missing and Invisible Energy}
\author{Mat\'ias Tueros}
\address{IFLP - CCT La Plata - CONICET, Casilla de Correo 727 (1900) La Plata, Argentina}                                                                                         
\address{Depto. de Fisica, Fac. de Cs. Ex., Universidad Nacional de La Plata,  Casilla de Coreo 67 (1900) La Plata, Argentina}
\ead{tueros@fisica.unlp.edu.ar}
%---------abstract----
\begin{abstract}

The energy carried away by neutral particles in ultra high energy cosmic ray showers can not be detected by fluorescence detectors. This energy is usually referred to as the \textit{invisible energy}. Since every shower has a fraction of invisible energy, the energy determined using the fluorescence technique is always less than the primary energy and a correction needs to be applied. This correction, usually referred to as the \textit{missing energy}, can only be estimated using Monte Carlo simulations. 

In this article we study in detail the influence that discarding low energy particles from the simulation has on the estimation of the missing and invisible energies. We found that although the effect is not important for the invisible energy,an important bias on the missing energy is introduced that can reach 30\% or more depending on the low energy cut value. 

We present a prescription on how to correct for this bias in AIRES simulations and give a novel missing energy parametrization including results for photons and for the QGSJET-II hadronic model. We also show that although missing and invisible energies are closely related they are conceptually different ideas if we consider the medium contribution to the shower energy.

\end{abstract}
%----------PACS--
%----------PACS--
\vspace{2pc}
\submitto{\JPG}
\maketitle
\section{Introduction}
\label{sec:Intro}
%The Importance of the invisible energy in the Fluorescence Technique.
%Limitations of simulations, low energy cut.
When a Ultra High Energy Cosmic Ray (UHECR) hits the atmosphere a cascade of particles is generated and an important but incomplete fraction of its energy is deposited through the electromagnetic channel as ionization of the air molecules and atoms.

A tiny, known fraction of the total deposited energy is re emitted as fluorescence light that can then be detected by ground telescopes. This phenomenon is the basis for the cosmic ray fluorescence detection technique pioneered by the Fly's Eye experiment \cite{Baltrusaitis} and implemented in Hi-Res \cite{Abu} and the Auger Observatory \cite{Engeniering} to determine the energy of cosmic rays.

In this technique, the primary \textit{calorimetric} energy is determined integrating the shower energy deposit profile in the atmosphere. Usually only a fraction of the longitudinal development of the shower is in the detector field of view, but the complete profile is obtained using an extrapolation of a Gaisser-Hillas function fit to the measured data.

The energy carried away by neutral particles in the shower can not be detected by fluorescence detectors as neutral particles do not produce ionization. This energy is usually referred to as the \textit{invisible energy}. Since there is always a fraction of invisible energy, the measured calorimetric energy is always less than the primary energy and to obtain the primary energy a correction needs to be applied.

This correction, usually referred to as the \textit{missing energy}, is known to be a function of the total deposited energy, the primary particle, and the hadronic model in use \cite{Linsley} \cite{Baltrusaitis0} \cite{Song} \cite{Barbosa}. This dependences make the missing energy one of the main contributions to the systematic uncertainties of the fluorescence method for the determination of the primary energy.

Note that although the missing and the invisible energy are closely related, they are not equivalent. The invisible energy is the energy in the shower not producing ionization. The missing energy is the difference between the primary energy and the calorimetric energy. The missing and the invisible energy would coincide only if there were no contributions from the medium to the shower energy pool. 

As it is unpractical (if not unfeasible when nuclear and electronic capture effects at low energies are considered) to follow all the particles to their rest in UHECR simulations, particles with energy falling below a given cut value ($E_{Cut}$) are not tracked by the simulation program and the fate of the energy they carry is not determined. Part of this energy might end deposited in the atmosphere and would thus contribute to the total emitted light, and part might go to neutrinos or other neutral particles, contributing to the invisible energy of the shower. 

We have shown elsewhere that this artificial cut in the simulation introduces a bias on all the observables related to the energy, like the total energy deposited, the total electromagnetic energy and the mean energy deposit per particle \cite{Tueros} and so it does on the missing and the invisible energy.

In this article we present a prescription to correct the introduced bias in the estimation of the missing and the invisible energy in simulations made with AIRES\cite{AiresManual}, complementing the prescription we presented in \cite{Tueros} for the energy deposit. Using a $2\times10^{4}$ shower library described in section \ref{sec:Simulations} the prescription and its effects are presented in section  \ref{sec:InvisibleE}. As an application of this prescription we present in section \ref{sec:MissingE} time a missing energy correction parametrization made with AIRES, that include novel results for photons and the QGSJET-II model.

The low energy cut is introduced to reduce the required CPU time in the simulations. Increasing $E_{Cut}$ reduces the amount of CPU time per shower at the expense of increasing the total amount of energy discarded from the simulation, making more important the necessary correction. We address the influence of $E_{Cut}$ on the calorimetric and invisible energy in section \ref{sec:InfluenceEnergyThreshold}.

\section{About the simulations}
\label{sec:Simulations}

The quantitative results presented in this work are based on the same set of AIRES simulations we used in \cite{Tueros}. AIRES  has been extensively used by many scientist around the world for the past ten years, and has become one of the standard simulation codes in the field \cite{AiresManual}.

AIRES includes the most important processes that may undergo shower particles from a probabilistic point of view. For the estimation of the shower invisible energy, neutron and neutrino production plays a central role. Neutrons are generated in the hadronic core of the shower and in photo-nuclear reactions. Their generation and propagation is treated through the external hadronic models QGSJET \cite{QGSJET},QGSJET-II \cite{QGSJETII} and Sybill \cite{Sybill} at high energies, and the EHSA \cite{EHSA} at low energies. Neutrinos are generated in decays of unstable particles (specially muons and pions) and are accounted for their number and energy, but they are not propagated.

The shower library used in this work was generated at the in2p3 computing centre \cite{Lyon} and has the following characteristics:

\begin{small}
Series name: AMgeLyonExtDvezpShb, generated by Sergio Sciutto\\
Hadronic models: QGSJET-II and SIBYLL\\
Primary particles: Proton, Iron and Photon (With MAGICS Preshower)\\
Ergy ($Log_{10}(eV)$) : 17.5, 18, 18.5, 19, 19.25, 19.5, 19.75, 20, 20.25, 20.5\\
Zenith (deg):0, 18, 25, 32, 36, 41, 45, 49, 53, 57, 60, 63, 66, 70, 72, 75, 78, 81, 84, 87\\
Thinning Energy Rel.: 1.0E-06     Thinning W. Factor: 0.1 \\
Gamma Cut Energy = 0.9 MeV\\
Electron/Positron Cut Energy = 0.4 MeV\\
MuonCutEnergy = 2.5 MeV\\
MesonCutEnergy = 4.5 MeV\\
NuclCutEnergy = 95 MeV\\
Thinning refers to a statistical sampling procedure to reduce computing time by simulating only a representative sample of all the particles. Thinning and thinning related parameters are explained in detail in the AIRES user manual \cite{AiresManual}.\\
\end{small}

In AIRES, different $E_{Cut}$ values can be set for different particle species. The ones more important for the study made in this work are electron/positron and gammas $E_{Cut}$ values, and we will always be referring to this particular cut throughout the text unless explicitly stated otherwise. Note that in this library $E_{Cut}$ values for gamma and electron/positron are low enough to determine unambiguously the fate of the particles. Gamma below 900 KeV wont be able to generate pairs, and will deposit all their energy. Electrons generated will also deposit all their energy until captured, and positrons will annihilate producing 2 or 3 gammas with energy below 1 MeV and thus unable to generate new pairs.

\section{Treatment of discarded and ground particles}
\label{sec:InvisibleE}
The correct treatment of discarded low energy particles regarding the shower energy balance and the total atmospheric energy deposit in AIRES has been addressed in \cite{Tueros}, using considerations similar to those presented in \cite{Barbosa} and \cite{Risse} for CORSIKA simulations. In that article, we proposed a prescription to treat the energy deposit of the discarded low energy particles ($E_{Dis}^{Deposit}$):

\begin{eqnarray} \label{eq:EDisDeposit}
E_{Dis}^{Deposit} & = & 0.997E_{Dis\:\gamma}+0.997E_{Dis\:\pi^{0}}+0.998E_{Dis\:e^{+/-}} \nonumber\\
                  &   & {} + 0.425E_{Dis\:\mu^{+/-}}+0.46E_{Dis\:\pi^{+/-}}
\end{eqnarray}

where $E_{Dis\:x}$ is the kinetic energy carried by the discarded low energy \textit{x} particles, easily available in AIRES from output tables 7501 trough 7892. The numerical factors where taken from GEANT4 simulations giving the average "releasable energy fraction" for each particle species.

The energy deposit from the rest mass of unstable discarded low energy particles ($E_{Dis\:rest\:mass}^{Deposit}$) is also included in the prescription as:

\begin{eqnarray} \label{eq:EDisRestMass}
E_{Dis\:rest\:mass}^{Deposit} & = & \frac{1}{3}m_{\mu}N_{Dis\:\mu}+\frac{1}{3}m_{\pi}N_{Dis\:\pi}+2m_{e}N_{Dis\:e^{+}}
\end{eqnarray}

where  $N_{Dis\:x}$ is the number of discarded \textit{x} particles, taken from AIRES output tables 7001 through 7293. The numerical factors come from the assumption that particles more likely to decay (muons and pions) deposit only 1/3 of their rest mass, and the remaining energy is carried away by the resulting neutrinos. Particles most likely to annihilate (positrons) contribute with twice their rest mass to the shower energy, taking its annihilation partner from the medium. AIRES classifies discarded low energy particles only for electrons, muons and gammas. Other particles are merged together in "other charged" and "other neutral" classes that in this work are all assumed to be pions.

For the invisible energy, we propose in this article the complementary equations. The invisible fraction of the discarded low energy particles kinetic energy ($E_{Dis}^{Invisible}$) becomes: 

\begin{eqnarray} \label{eq:EDisInv}
E_{Dis}^{Invisible} & = & 0.003E_{Dis\:\gamma} + 0.003E_{Dis\:\pi_{0}} + 0.002E_{Dis\:e^{+/-}}  \nonumber \\
                    &   & {} + 0.575E_{Dis\:\mu} + 0.54E_{Dis\:\pi^{+/-}}
\end{eqnarray}

and the invisible fraction of the discarded particles rest mass ($E_{Dis\:rest\:mass}^{Inv}$):

\begin{eqnarray} \label{eq:EDisInvRestMass}
E_{Dis\:rest\:mass}^{Inv}=\frac{2}{3}m_{\mu}N_{Dis\:\mu}+\frac{2}{3}m_{\pi}N_{Dis\:\pi}
\end{eqnarray}

In the usual method for the estimation of the primary electromagnetic energy of a cosmic ray shower using the fluorescence technique \cite{Abu}\cite{Song}, the measured light profile is fitted with a Gaisser-Hillas function that is then extrapolated to get the complete light profile, since it is rare the occasion where the complete shower develops completely in the detector field of view or before reaching ground level. Using the photon yield per deposited MeV and the corresponding geometrical, detector and atmospheric related correction factors, the extrapolated light profile is integrated to get the \textit{total energy deposit the shower would have had} if it had developed completely. This is usually referred to as the "calorimetric" energy of the shower, that needs then to be corrected for the missing energy to get the primary energy.

As the calorimetric energy is calculated on the \textit{complete} shower profile extrapolated from a fit of the Gaisser-Hillas function to the measured profile, we have to include in the analysis of our simulation the energy of the shower particles reaching ground level. If the shower had continued developing further, part of this energy would have been deposited completing the light deposit profile and part would have been lost to invisible channels. 

The fraction of energy that would go to ionization and the fraction that would be invisible energy is a priori unknown and we use the results from \cite{Barbosa} to estimate it. For species not contemplated in \cite{Barbosa}, an arbitrary factor of 0.5 was chosen.
\begin{eqnarray} \label{eq:EGroundDeposit}
E_{Ground}^{Deposit} & = & 0.997E_{Ground\:\gamma}+0.998E_{Ground\:e^{+/-}}+0.593E_{Ground\:\pi^{+}} \nonumber \\
                     &   & {} +0.617E_{Ground\:\pi^{-}}+0.604E_{Ground\:K}+0.753E_{Ground\:p} \nonumber \\
                     &   & {} +0.732E_{Ground\:pbar}+0.701E_{Ground\:N}+0.5E_{Ground\:other}
\end{eqnarray}
\begin{eqnarray} \label{eq:EGroundInvisible} 
E_{Ground}^{Invisible} & = & 0.003E_{Ground\:\gamma}+0.002E_{Ground\:e^{+/-}}+1.0E_{Ground\:\mu} \nonumber \\
                       &   & {} +0.407E_{Ground\:\pi^{+}}+0.383E_{Ground\:\pi^{-}}+0.396E_{Ground\:K} \nonumber \\
                       &   & {} +0.247E_{Ground\:p}+0.268E_{Ground\:pbar}+0.299E_{Ground\:N}\nonumber\\
                       &   & {} +0.5E_{Ground\:other}
\end{eqnarray}
Note that in these equations the kinetic energy of muons arriving to the ground is considered as invisible energy. Although muons do produce ionization their cross section is so small that their longitudinal development is not described by the Gaisser-Hillas function fitted to the fluorescence shower data, dominated by the electrons and gammas. Although muons energy is theoretically visible for the fluorescence detectors, no shower will develop enough in the atmosphere to deposit all the muons energy, making it invisible in practice.

The rest mass of ground particles must also be accounted for and we follow the decay corrections guidelines from \cite{Risse} to estimate it. As the vast majority of the electrons arriving at ground level are taken from the air in knock-on collisions, the rest mass of ground electrons is ignored. Particles more likely to annihilate (positrons and pbars) contribute with twice their rest mass, as their add the antiparticle rest mass to the shower energy pool.
\begin{eqnarray}\label{eq:DepositGroundRestMass}
E_{Ground\:rest\: mass}^{Deposit} & = & 2m_{e^{+}}N_{Ground\:e^{+}}+\frac{1}{3}m_{\mu}N_{Ground\:\mu}+\frac{1}{3}m_{\pi}N_{Ground\:\pi} \nonumber \\
                                  &   &{} +\frac{1}{3}m_{K}N_{Ground\:K}+m_{p}N_{Ground\:p}+2m_{p}N_{Ground\:pbar}\nonumber\\
                                  &   &{} +\frac{3}{4}m_{N}N_{Ground\:N}
\end{eqnarray}
\begin{eqnarray}\label{eq:InvisibleGroundRestMass}
E_{Ground\:rest\:mass}^{Invisible} & = & \frac{2}{3}m_{\mu}N_{Ground\:\mu}+\frac{2}{3}m_{\pi}N_{Ground\:\pi} \nonumber \\
                                   &   & {} +\frac{2}{3}m_{K}N_{Ground\:K}+\frac{1}{4}m_{N}N_{Ground\:N}
\end{eqnarray}
The information on the number of particles reaching ground level and their total energy is available in the GRD record of longitudinal profile tables 1001 to 1293 and 1501 to 1793.

Using equations (\ref{eq:EDisDeposit}) to (\ref{eq:InvisibleGroundRestMass}), the total shower energy can be divided in two different components: calorimetric energy and invisible energy.

\subsection{Calorimetric Energy}

The Calorimetric Energy ($E_{Cal}$) is the energy that was deposited or would have been deposited, capable of producing ionization and fluorescence light.
\begin{equation} \label{eq:ECal}
E_{Cal}=E_{Dep} +  E_{Dis}^{Deposit} + E_{Dis\:rest\:mass}^{Deposit} + E_{Ground}^{Deposit} + E_{Ground\:rest\:mass}^{Deposit}
\end{equation}
where

\begin{itemize}[itemsep=10pt]
\item $E_{Dep}$:Energy deposit in the atmosphere in the simulation.
\item $E_{Dis}^{Deposit}$: Energy deposit from discarded particles kinetic energy.
\item $E_{Dis\:rest\:mass}^{Deposit}$: Energy deposit from discarded particles rest mass.
\item $E_{Ground}^{Deposit}$:Depositable fraction of ground particles kinetic energy.
\item $E_{Ground\:rest\: mass}^{Deposit}$: Depositable fraction of ground particles rest mass.
\end{itemize}

\begin{figure} %this plot was generated using paperviewplots.c
\begin{center} 
\includegraphics[angle=0,scale=0.7]{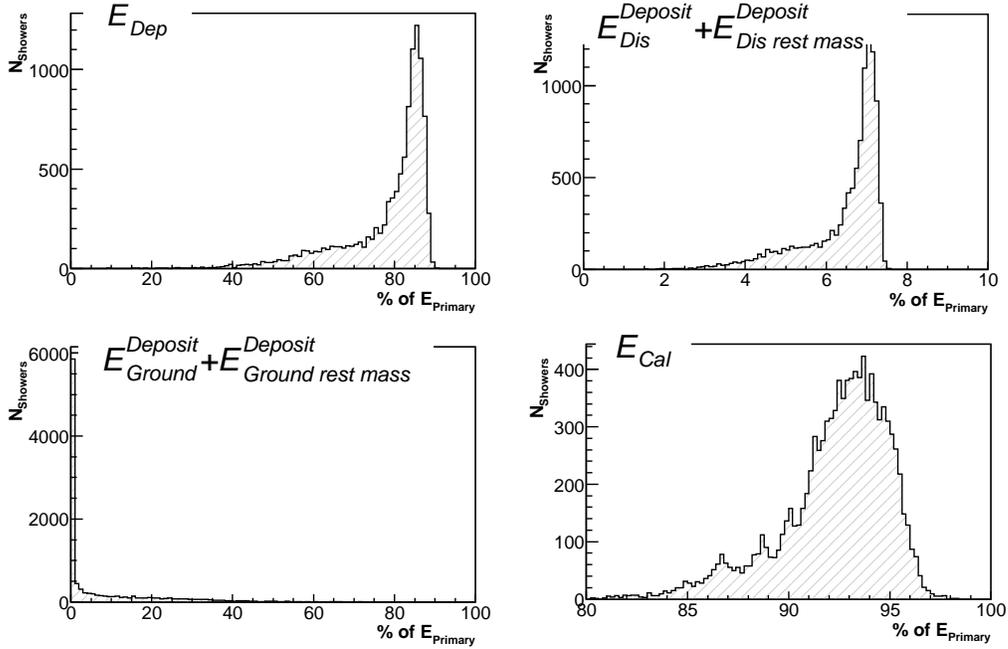} 
\caption{Histograms showing the total calorimetric energy ($E_{Cal}$) for Proton and Iron primaries and the discarded and ground particles contributions to it as a percentage of the primary energy. $E_{Dep}$:Energy deposit in thee atmosphere in the simulation. $E_{Dis}^{Deposit}$: Energy deposit from discarded particles kinetic energy.$E_{Dis\:rest\:mass}^{Deposit}$: Energy deposit from discarded particles rest mass. $E_{Ground}^{Deposit}$:Depositable fraction of ground particles kinetic energy.$E_{Ground\:rest\: mass}^{Deposit}$: Depositable fraction of ground particles rest mass.}
\label{gr:EDep}
\end{center}
\end{figure}

When all the terms are added together it can be seen that the total calorimetric energy represents between 80 and 97\% of the primary energy for proton and iron showers, depending mainly on the shower energy as it will be shown in section \ref{sec:MissingE}. 

The contribution of discarded particles to the calorimetric shower varies with shower age at ground level. Showers with a lower primary energy or a a higher zenith angle will have developed more in the atmosphere, depositing more energy but also loosing more particles to the low energy cut and having less available energy at ground level. High energy, vertical showers can't even reach shower maximum before reaching ground level, and a lot of energy is still left in the particles at ground level while little energy was deposited or lost by the low energy cut. This gives a very large spread in $E_{Dep}$, $E_{Dis}$ and $E_{Ground}$ terms in (\ref{eq:ECal}), as can be seen in figure \ref{gr:EDep}.

For photons, the calorimetric energy distribution is very sharp peaked around 99\% of the primary energy, and virtually independent of the simulation details, as seen in figure \ref{gr:Photons}.

Discarded low energy particles carry about 7 \% of the primary energy in our simulation sample, representing a 8\% correction to $E_{Cal}$ but the contribution will increase on simulations made with higher $E_{Cut}$ values, as will be shown on section \ref{sec:InfluenceEnergyThreshold}.

\subsection{Invisible Energy}

The invisible energy component is the amount of energy not capable of producing ionization and fluorescence light
\begin{equation} \label{eq:EInv}
E_{Invisible}=E_{\nu} +  E_{Dis}^{Invisible} + E_{Dis\:rest\:mass}^{Invisible} + E_{Ground}^{Invisible} + E_{Ground\:rest\:mass}^{Invisible}
\end{equation}
where
\begin{itemize}[itemsep=10pt]
\item $E_{\nu}$:Energy of neutrinos.
\item $E_{Dis}^{Invisible}$: Discarded particles invisible kinetic energy.
\item $E_{Dis\:rest\:mass}^{Invisible}$: Invisible Energy from discarded particles rest mass.
\item $E_{Ground}^{Invisible}$: Invisible kinetic energy of particles reaching ground.
\item $E_{Ground\:rest\:mass}^{Invisible}$: Invisible energy from the rest mass of particles reaching ground.
\end{itemize}

\begin{figure} %this plot was generated using paperviewplots.c
\begin{center} 
\includegraphics[angle=0,scale=0.7]{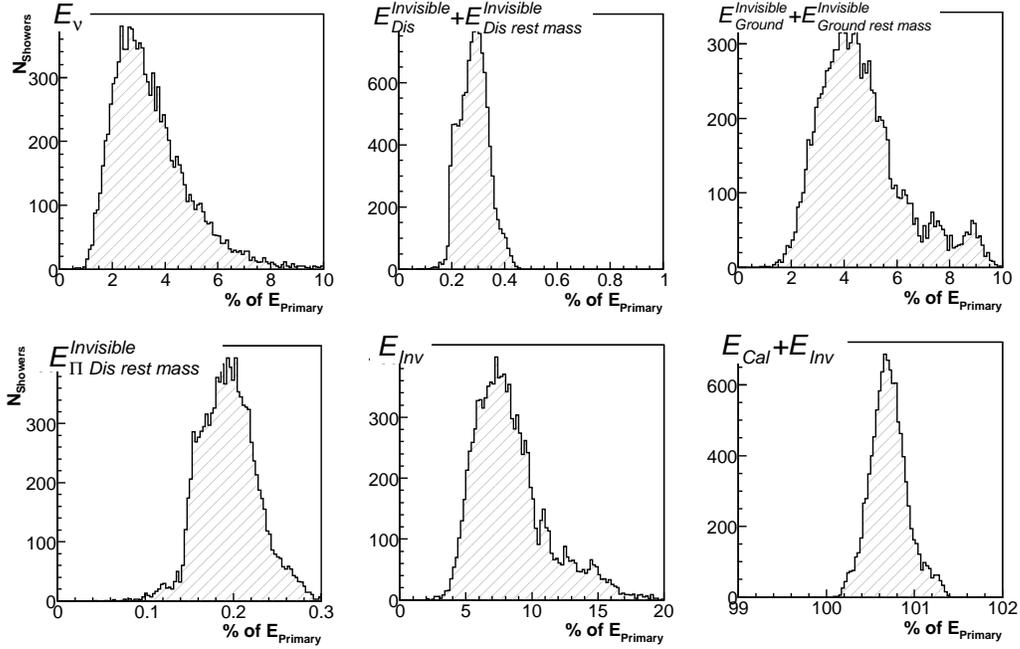} 
\caption{Histograms showing the total invisible energy ($E_{Inv}$) for Proton and Iron primaries and the contribution from discarded and ground particles to it as a percentage of the primary energy. $E_{\nu}$: Energy of neutrinos. $E_{Dis}^{Invisible}$: Discarded particles invisible kinetic energy. $E_{Dis\:rest\:mass}^{Invisible}$: Invisible Energy from discarded particles rest mass. $E_{Ground}^{Invisible}$: Invisible kinetic energy of particles reaching ground. $E_{Ground\:rest\:mass}^{Invisible}$: Invisible energy from the rest mass of particles reaching ground. The total Energy in the shower, $Total\:E$ is also shown.}
\label{gr:EInv}
\end{center}
\end{figure}

\begin{figure} %this plot was generated using paperviewplots.c
\begin{center} 
\includegraphics[angle=0,scale=0.7]{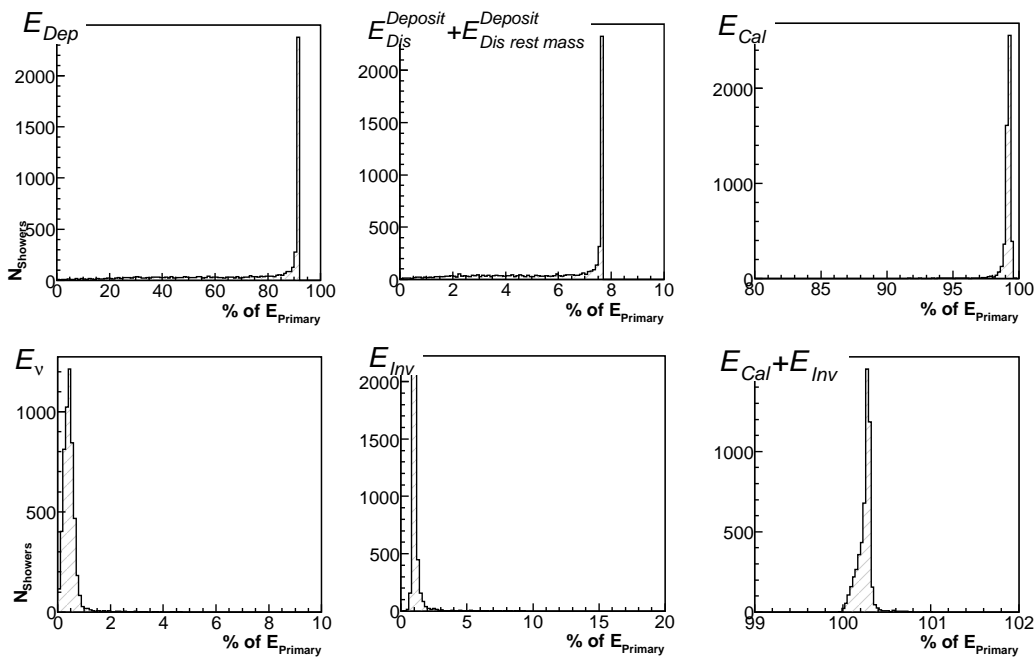}
\caption{Top row: Histograms showing the simulation energy deposit, the contribution from discarded particles and the total energy deposit for photon primaries. Bottom row: Histograms showing the invisible energy carried away by neutrinos, the total invisible energy and the total shower energy for photon primaries as a percentage of the primary energy. These histograms have the same scale as their counterparts in figures 1 and 2 to make comparisons easier.}
\label{gr:Photons}
\end{center}
\end{figure}

For hadronic showers, neutrinos and ground energy account on average for more than 95-97 \% of the total invisible energy. Low energy particles contribute with the remaining 3-5\% that represents only a 0.4\% in terms of the primary energy, as shown on figure \ref{gr:EInv}. For photons, low energy particles are even less important since the bulk of the invisible energy is provided by neutrinos (figure \ref{gr:Photons}).

As the variation of the invisible energy with other shower parameters like primary energy, primary mass or hadronic model is in the range of several \% of the primary energy, the low energy particles correction to the invisible energy is negligible. Making a detailed study to determine the arbitrary factors used in (\ref{eq:InvisibleGroundRestMass}) and (\ref{eq:DepositGroundRestMass}) seems unjustified.

The total invisible energy is insensitive, within reasonable limits, to the $E_{Cut}$ value used in the simulation. The main source of invisible energy is the hadronic core of the shower where high energy collisions generating neutrinos, muon and pions take place. The invisible energy taken away by discarded low energy muons, pions and neutrons is in comparison very small, only 0.6 \% of the primary energy (figure \ref{gr:EInv}). We will comment further on this in section \ref{sec:InfluenceEnergyThreshold}.

\subsection{Medium Energy}

The bottom right panels of figures \ref{gr:EInv} and \ref{gr:Photons} show that the sum of $E_{Cal}$ and $E_{Invisible}$ accounts for more than the primary energy. As the shower develops in the atmosphere, energy from the rest mass of air nuclei is added to the shower energy pool in every nuclear collision. Antiparticles generated in the shower (i.e. anti-protons) also add energy from the medium to the shower when they annihilate. This extra energy is 0.8 $\pm$ 0.6 \% of the primary energy in hadronic showers, and about a factor four less in photon showers.

Note that although we expect all the showers to have more than the primary energy, there are still some photon showers with less than the primary energy. This comes from the fact that in equations (\ref{eq:EDisDeposit}) through (\ref{eq:EDisInvRestMass}) "other particles" where assumed to be pions, providing a lower limit. Particles with higher mass are surely present. Figure \ref{gr:EInv} shows the contribution from this assumed pions is around 0.3\% ($0.2\%\times\frac{3}{2}$, see (\ref{eq:EDisInvRestMass})). 

\begin{figure} %this plot was generated using paperviewplots.c
\begin{center} 
\includegraphics[angle=0,scale=0.7]{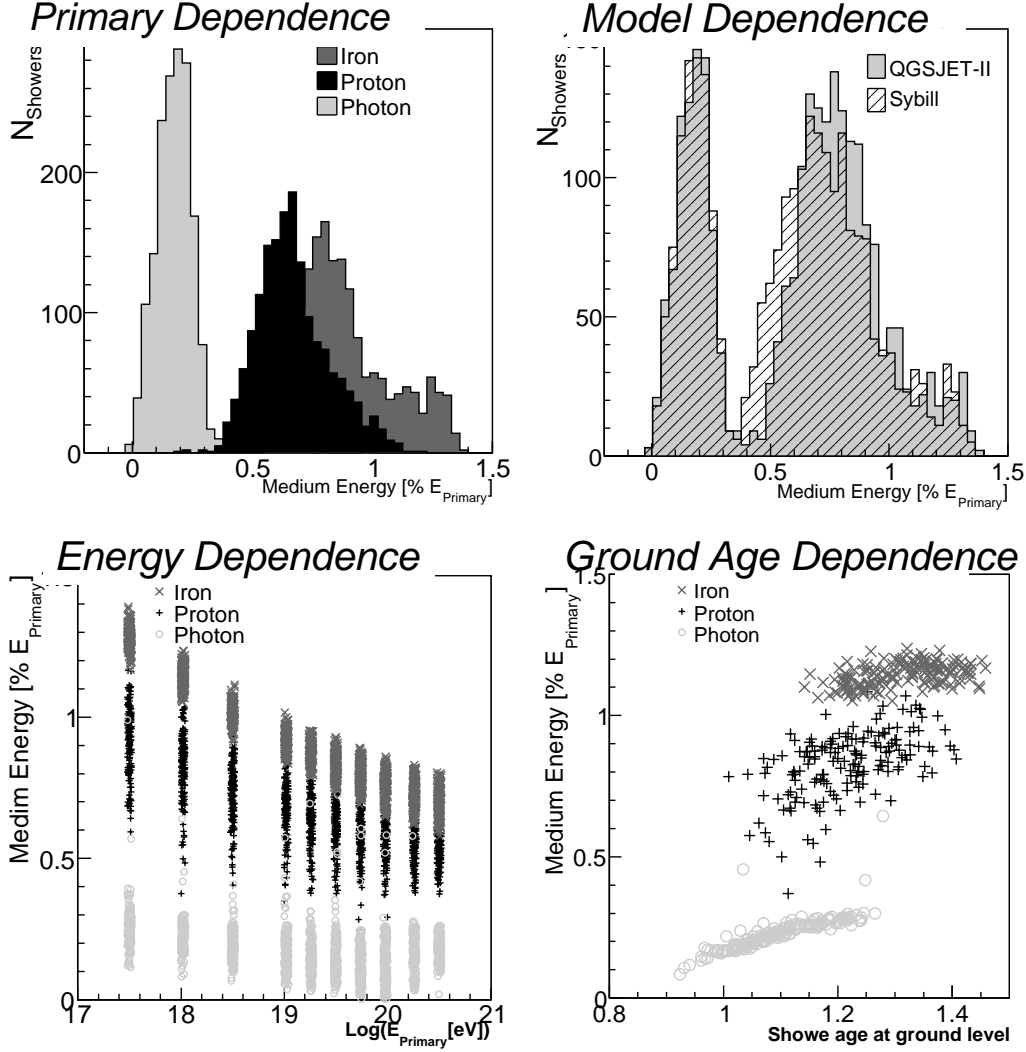}
\caption{Primary, model, energy and ground age dependences of the medium contribution to the shower energy.}
\label{gr:Medium}
\end{center}
\end{figure}

If "other particles" are considered nucleons to give an upper limit, the contribution reaches 1.7\% shifting the total energy and the medium contribution an average 1.4 \%, bringing it close to 2.2\%. The real value is of course somewhere in between. Treating in detail medium contributions to the shower energy is beyond the scope of this article,but some interesting trends are shown in figure \ref{gr:Medium}. A detailed study is left for a future paper.

\section{Missing Energy Correction}
\label{sec:MissingE}
We have seen in the previous section that the calorimetric energy does not account for all the primary energy. The difference between the primary energy and the calorimetric energy is usually referred to as the \textit{missing energy} and is directly related to the invisible energy of the shower but it is always lower, as we have shown that the total energy is always higher than the primary energy.

\begin{equation}
 E_{Missing}=E_{Primary}-E_{Cal} < E_{Invisible}=E_{Total}-E_{Cal}
\label{eq:EMissing}
\end{equation}

The main sources of invisible energy during the shower development are neutrino, muon and pion production making the missing energy highly dependent on the hadronic model and on the primary mass. As the mass composition of the primary particle is \textit{a priori} unknown, an average correction between Proton and Iron is made and the spread between these two corrections is one of the main systematic uncertainties in the fluorescence method.

For a given primary and hadronic model, a parametrization of  $\frac{E_{Cal}}{E_{Primary}}$ vs $E_{Cal}$ is used to estimate the total energy \cite{Barbosa}
\begin{equation}\label{eq:parametriz}
\frac{E_{Cal}}{E_{Primary}}=\alpha - \beta\times(\frac{E_{Cal}}{1EeV})^{-\gamma}
\end{equation}

\begin{figure} %this plot was generated using viewplots.c
\begin{center} 
\includegraphics[angle=0,scale=0.7]{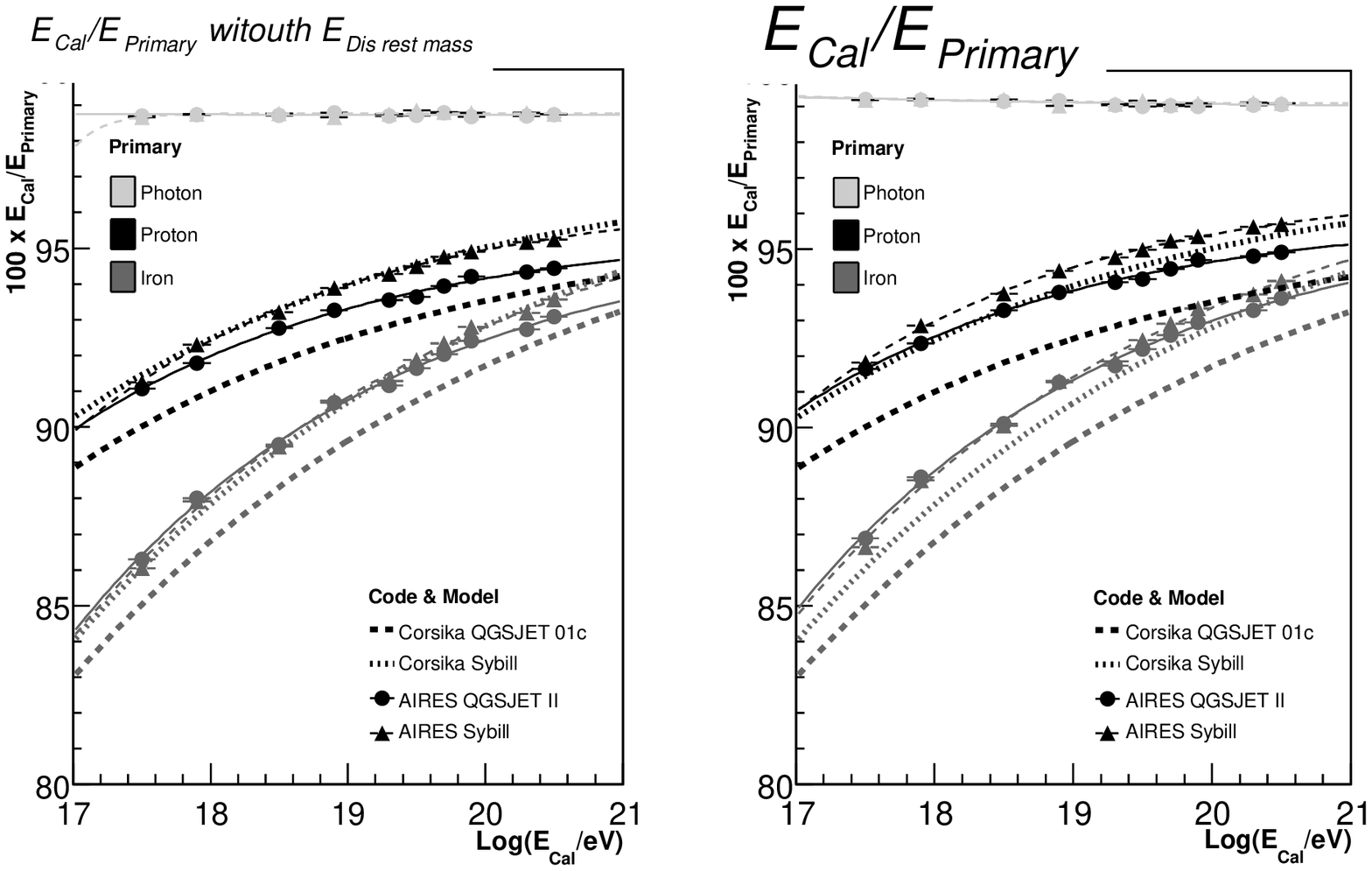}
\caption{Missing energy correction for proton, iron and photon primaries. The left panel shows the correction without considering the discarded particles rest mass to enable comparisons with CORSIKA results from \cite{Barbosa}. The right panel shows the complete correction.}
\label{gr:MissParam}
\end{center}
\end{figure}

The parametrization values found with AIRES including the corrections presented in this article for different Hadronic models and primary particles are given in Table \ref{tb:Miss} and is compared with the results obtained in \cite{Barbosa} for CORSIKA simulations in the right side of figure \ref{gr:MissParam}. Comparing both simulations codes using the same model (Sybill) we can see that the differences are below 1\% of the primary energy. 

It is interesting to note that if we exclude from our analysis the discarded particles rest mass, we get a much better agreement (left panel of figure \ref{gr:MissParam}). This component must have been ignored in \cite{Barbosa}, as no reference to the discarded low energy particles rest mass is made on it.

\begin{table}
\caption{\label{tb:Miss}Parameters for the missing energy parametrization (\ref{eq:parametriz}) for different primaries and hadronic models.}
\begin{indented}
\lineup 
\item[]\begin{tabular}{@{}lllll}
\br
Model & Primary & $\alpha$ & $\beta$ & $\gamma$  \\ 
\mr
Sybill & Photon & 0.990 & \-0.00132 & 0.2401 \\ 
Sybill & Proton & 0.968 & 0.03804 & 0.2223 \\ 
Sybill & Iron & 0.978 & 0.09180  & 0.1557 \\
QGSJETII & Photon & 0.989 & \-0.00261 & 0.1213 \\   
QGSJETII & Proton & 0.959 & 0.03402 & 0.2077 \\ 
QGSJETII & Iron & 0.960 & 0.07271  & 0.1875 \\ 
\br
\end{tabular} 
\end{indented}
\end{table}

The missing energy defined in (\ref{eq:EMissing}) depends strongly on $E_{Cut}$ through $E_{Cal}$, that has been show to be very sensitive to it in \cite{Tueros}. We will show in the next section that the corrections presented in this article correctly account for this effect.

For studies focusing only on the missing energy, high $E_{Cut}$ values can be used in the simulations to save CPU time. In those cases, the invisible energy should be used as a substitute, as it is very similar to the missing energy for air showers and is independent of $E_{Cut}$.

It was shown in \cite{Tueros} that setting $E_{Cut}$ to 100 MeV reduces the required CPU time almost 10 fold when compared to a standard 1 MeV $E_{Cut}$ and 20 fold when compared to a more detailed 0.1 MeV $E_{Cut}$. Going to $E_{Cut}$ values higher than 210 MeV however for electron and gammas is not recommended, as gammas producing muon pairs that would contribute to the invisible energy will not be simulated.

\section{Dependency with the Low Energy Cut value}
\label{sec:InfluenceEnergyThreshold}
The algorithms presented in this article to estimate the contribution of the discarded particles are independent of the $E_{Cut}$ value, and have been tested up to 100 MeV $E_{Cut}$.

\begin{figure} %this plot was generated using viewplotsecut.c
\begin{center} 
\includegraphics[angle=0,scale=0.7]{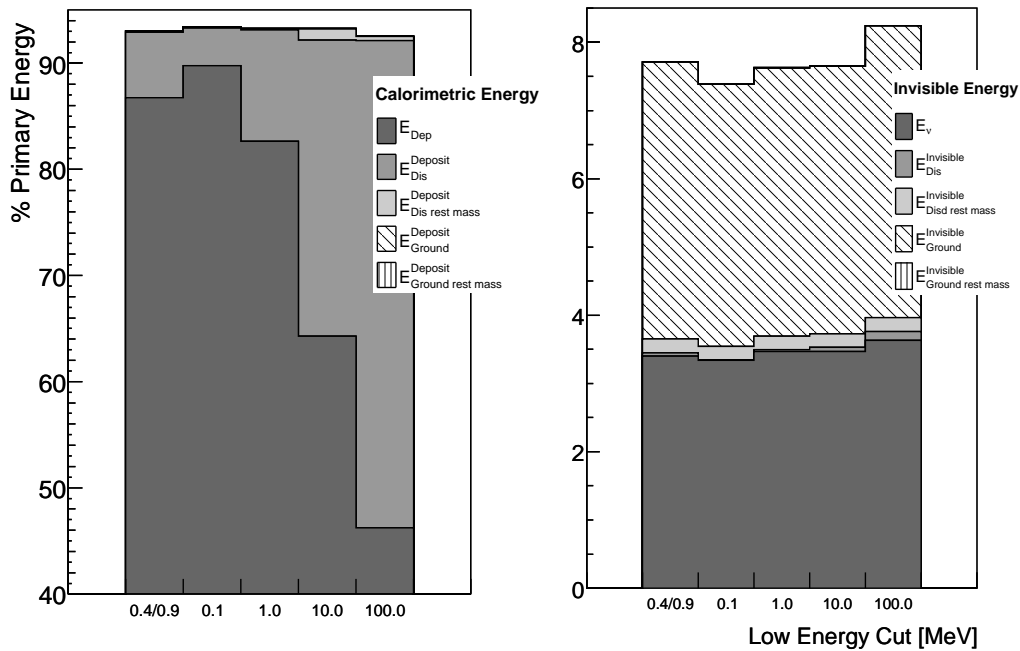}
\caption{Components of the Calorimetric energy (left) and of the Invisible Energy (right) of 1 EeV, 60 deg Zenith proton showers for different energy cuts. The total energy remains constant within shower to shower fluctuations, while deposited energy is traded for discarded particles energy as the energy cut rises. The correction to the total Calorimetric energy from discarded low energy particles kinetic and rest mass energy is very important and can reach up to 50\% of the primary particle energy, while the Invisible Energy, originated in the high energy interactions in the shower has little dependence on the low energy cut.}
\label{gr:Ecut}
\end{center}
\end{figure}
 
The higher the energy cut the higher the amount of discarded energy, and the impact of the corrections presented on the previous sections on the shower observables.

Figure \ref{gr:Ecut} (right) illustrates how the energy deposit on the simulation is lowered as more and more energy is lost to the low energy particle cut. If we consider the energy deposit and the corrections proposed by (\ref{eq:EDisDeposit}), (\ref{eq:EDisRestMass}), (\ref{eq:EGroundDeposit}) and (\ref{eq:DepositGroundRestMass}) we see that the total calorimetric energy remains independent of the energy cut within shower to shower fluctuations. As $E_{Cut}$ is raised, the amount of discarded energy varies accordingly but the sum of the various energy terms of (\ref{eq:ECal}) remains constant within shower to shower fluctuations.

Figure \ref{gr:Ecut} is shown as a particular example and uses 1 EeV proton showers with 60 deg zenith. These showers develop almost completely in the atmosphere and arrive at ground level with little electromagnetic energy, making irrelevant the contribution from ground particles to $E_{Cal}$. The left side of figure \ref{gr:Ecut} however shows that ground reaching particles (most of them muons) do have an important contribution to the invisible energy. The total invisible energy, as mentioned in previous sections, remains insensitive to $E_{Cut}$, since its main dependence is on the hadronic model and primary mass.

The stability of the totals with $E_{Cut}$ shown in this figure is representative of all the showers in our simulations. Such stability provides confidence in the proposed algorithms to account for the discarded particles depositable and invisible energy at any energy and enables their use to make comparisons between simulations made with different $E_{Cut}$ values.

\section{Conclusions}

We have presented a prescription to estimate correctly the missing and the invisible energy in AIRES shower simulations, taking into account the energy carried away by the low energy particles discarded from the simulation. This prescription is independent of the $E_{Cut}$ value
and can be safely used up to a 210 MeV cut value, the muon pair production threshold.

We found that the total invisible energy is insensitive to the $E_{Cut}$ value used in the simulation, as its main source is at the core of the high energy collisions generating neutrinos, muon and pions in the shower. For simulations that end at ground level, it is mandatory to include the contribution that ground particles would have had if the shower had completed its development, as this can be well above half the total invisible energy.

For the missing energy, on the other hand, we found a strong dependence with $E_{Cut}$. The correction introduced is about an 8\% for the cut values used in this work but go to 10\% at 1 MeV, a 50\% at 10 MeV and a 100\% correction or more at 100 MeV $E_{Cut}$ values. We have shown that the proposed prescription correctly accounts for this effect, making it possible to compare results from simulations made with different $E_{Cut}$ values.

The difference on the impact that discarded low energy particles have on the invisible and missing energies is inherent to their definition, and puts in evidence that invisible and missing energies are conceptually different. Every time a nuclear interaction takes place the energy from the rest mass of the nucleon is added the shower energy pool, making invisible energy always higher than the missing energy and the total energy always higher than the primary energy.

As an application of this prescription we presented for the first time a parametrization of the missing energy correction made with AIRES simulations for proton, iron and photon primaries using the Sybill and QGSJET-II hadronic models. This parametrization can be used in the reconstruction of shower signals on fluorescence detectors to get the primary energy, once the calorimetric energy is known. 

\ack
We thank H. Wahlberg for his help running the analysis scripts in Lyon, and A. Mariazzi and S. Sciutto for useful comments and proof reading the manuscript. M. T. is supported by a grant of the Consejo Nacional de Investigaciones Científicas y Tecnológicas de la República Argentina.

%\listoffigures

\end{document}